\documentclass[10pt, twocolumn, aps, prl, superscriptaddress]{revtex4-2}
\usepackage[utf8]{inputenc}

\usepackage[dvipsnames]{xcolor}
\usepackage{amsmath}
\usepackage{amsfonts}
\usepackage{fullpage}
\usepackage{mathrsfs}
\usepackage{fancyhdr}
\usepackage{verbatim}
\usepackage{framed}
\usepackage{algorithmic}
\usepackage{enumerate}
\usepackage{hyperref}
\usepackage[capitalize]{cleveref}
\usepackage{bm}
\usepackage{amssymb}
\usepackage{amsthm}
\usepackage{mathtools}
\usepackage{cancel}
\usepackage{qcircuit}
\usepackage[explicit]{titlesec}
\usepackage{graphicx}
\usepackage{comment}
\usepackage{svg}

\titlespacing{\paragraph}{0pt}{0pt}{0pt}

\renewcommand{\section}[1]{\paragraph{#1---}}
\renewcommand{\subsection}[1]{\paragraph{#1---}}

\newcommand{\eps}{\varepsilon}
\newcommand{\clH}{\mathcal{H}}
\newcommand{\quH}{\hat{\mathbf{H}}}
\newcommand{\numax}{\nu_{\rm max}}

\newcommand{\ket}[1]{{|#1\rangle}}
\newcommand{\bra}[1]{{\langle#1|}}

\newcommand{\p}[1]{\left({#1}\right)}
\newcommand{\br}[1]{\left[{#1}\right]}

\definecolor{plum}{HTML}{5D5DA8}
\definecolor{evergreen}{HTML}{009447}
\definecolor{ruby}{HTML}{BC1D1F}

\renewcommand{\cal}[1]{\mathcal{#1}}

\DeclareMathOperator{\poly}{poly}

\newtheorem*{problem}{Problem}

\newif\ifnotes\notestrue
\ifnotes

\else
\newcommand{\znote}[1]{}
\newcommand{\anote}[1]{}
\newcommand{\tm}[1]{}
\fi

\def\vecx{\mathbf{x}}
\def\vecp{\mathbf{p}}

\def\vecy{\mathbf{y}}
\def\veck{\mathbf{k}}

\def\vecv{\mathbf{v}}
\def\matM{\mathbf{M}}

\def\matQ{\mathbf{Q}}
\def\matK{\mathbf{K}}
\def\matF{\mathbf{F}}
\def\matG{\mathbf{G}}
\def\matH{\mathbf{H}}
\def\matU{\mathbf{U}}

\def\matV{\mathbf{V}}
\def\matD{\mathbf{D}}
\def\matA{\mathbf{A}}
\def\matI{\mathbf{I}}

\def\matP{\mathbf{P}}
\def\matW{\mathbf{W}}
\def\matS{\mathbf{S}}

\def\matT{\mathbf{T}}

\def\cB{{\cal B}}

\def\cP{{\cal P}}
\def\cQ{{\cal Q}}

\begin{abstract}
    A recent promising arena for quantum advantage is simulating exponentially large classical systems. Here, we show how this advantage can be used to calculate the dynamics of open classical systems experiencing dissipation, including the effects of non-Markovian baths. This is a particularly interesting class of systems since dissipation plays a key role in contexts ranging from fluid dynamics to thermalization. We adopt the Caldeira-Leggett Hamiltonian, a generic model for dissipation in which the system is coupled to a bath of harmonic oscillators with a large number of degrees of freedom. To date, the most efficient classical algorithms for simulating such systems have a polynomial dependence on the size of the bath. In this work, we give a quantum algorithm with an exponential speedup, capable of simulating $d$ system degrees of freedom coupled to $N = 2^n\gg d$ bath degrees of freedom, to within error $\eps$, using $O(\poly(d, n, t, \eps^{-1}))$ quantum gates.
\end{abstract}

\begin{document}

\title{Exponential Quantum Advantage for Simulating Open Classical Systems}
\author{\'{A}gi Vill\'{a}nyi}
\email{agivilla@mit.edu}
\affiliation{Massachusetts Institute of Technology, 77 Massachusetts Ave, Cambridge, MA 02139, USA}
\affiliation{Laboratory for Physical Sciences, 8050 Greenmead Dr., College Park, MD 20740, USA}
\author{Yariv Yanay}
\email{yanay@umd.edu}
\affiliation{Laboratory for Physical Sciences, 8050 Greenmead Dr., College Park, MD 20740, USA}
\affiliation{Department of Physics, University of Maryland, College Park, MD 20742, USA}
\author{Ari Mizel}
\email{ari@lps.umd.edu}
\affiliation{Laboratory for Physical Sciences, 8050 Greenmead Dr., College Park, MD 20740, USA}
\maketitle

\section{Introduction}
The simulation of classical particle dynamics is a central problem in contexts as diverse as molecular dynamics, accelerator design, computer graphics, and food processing.  There is an established software industry that aims to address this computational need, but available computational resources inevitably place limits on the complexity of the simulations that can be performed. Recently, a clever quantum algorithm \cite{babbush-2023-oscillators} was proposed that offers an exponential speedup over classical computation for a specific dynamical system: an energy-conserving sparse network of classical harmonic oscillators. This algorithm offers a promising future application for quantum computing; to realize its potential, it is essential to extend the range of cases in which a quantum speedup is available.  
Here, we offer a key extension, using the algorithm to to simulate not an exponentially large system but rather an exponentially large bath generating non-conservative dynamics in finite system of interest. 

Modeling the dynamics of such open systems is of great importance in fields ranging from classical and fluid dynamics \cite{Nicolis1986,Espanol2017} to biological matter \cite{Marchetti2013,Goldbeter2018} to quantum mechanics \cite{Clerk2010}. In the realm of quantum computation and information, dissipative behavior is both a fundamental challenge in the form of decoherence \cite{Dube2001,Schlosshauer2019} and a powerful tool for stabilizing entangled states \cite{Verstraete2009,Yanay_2018,Harrington2022}.  Specifically, the simulation of non-Markovian systems, which describe a wide-range of physical systems from polymers to transport across biological membranes, poses a unique challenge due to the need to simulate an exponentially large bath \cite{van-1998-non-markov, Doi-1986-TheTO}. In this paper, we present an algorithm providing an exponential quantum speedup for open classical systems, including those experiencing non-Markovian effects.

\begin{figure}[t]
    \centering
    \includegraphics[width=\linewidth]{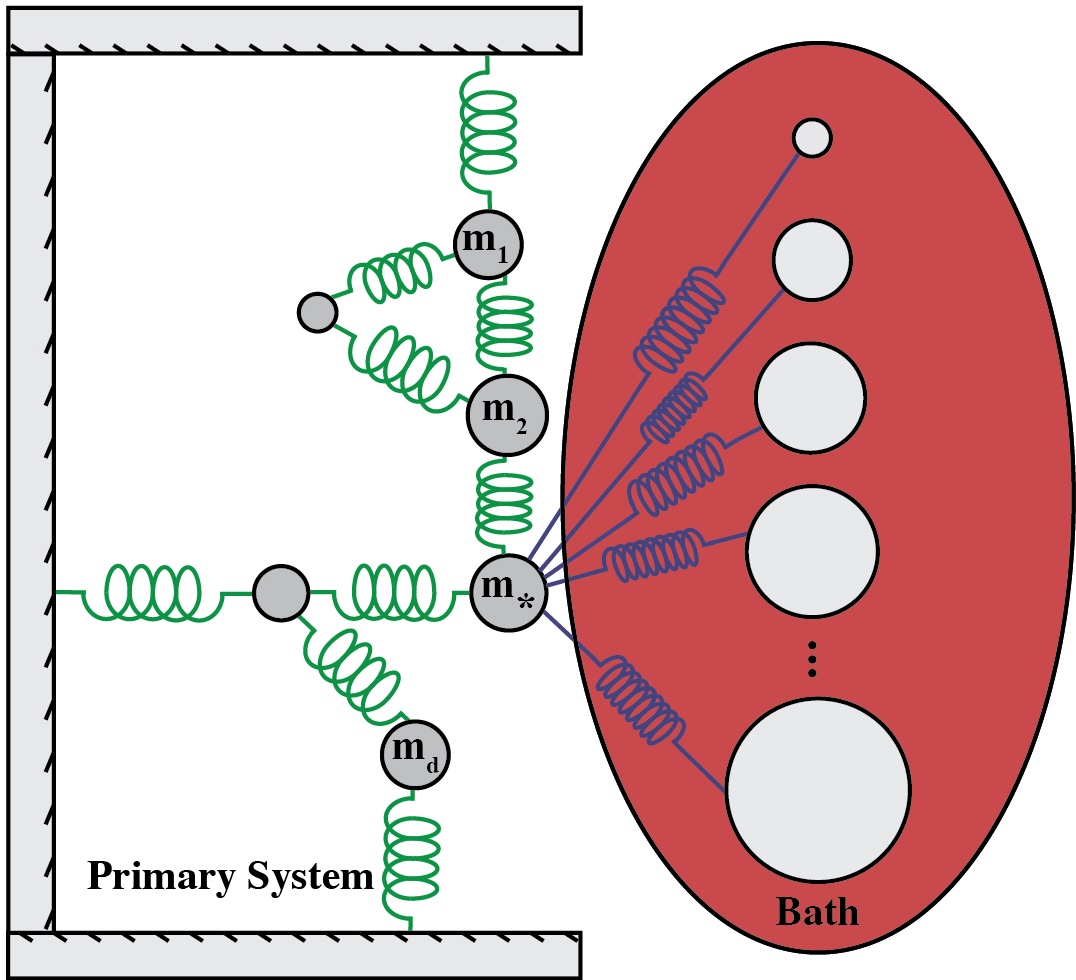}
    \caption{An example of a system coupled to a bath represented by the Caldeira-Leggett model. $d$ primary system degrees of freedom with mass $m_1, \dots, m_d$, are quadratically coupled to each other. A single mass $m_{*}$ is coupled to a dissipative bath,  modeled by a large number $N = 2^n\gg d$ of independent degrees of freedom with different frequencies.}
    \label{fig:system-fig}
\end{figure}

Our analysis employs the canonical Caldeira-Leggett (CL) model \cite{legget-1981-macroscopicTunneling, caldeira-1983-pathIntegralApproach, caldeira-1983-tunnellingDissipation},  which produces dissipative dynamics by representing the external bath as a large collection of harmonic oscillators (see \cref{fig:system-fig}).
We show how this bath can be represented as a Hamiltonian simulation problem, and how the resulting Hamiltonian can be accessed by employing non-sparse techniques for quantum simulation \cite{childs-2009-relationship, childs-2010-limitations, childs-2012-blackbox}. As a result, our algorithm is capable of capturing genuine non-Markovian effects in addition to the memory-less Markovian dissipation that arises in the limit of a uniform, infinite bath. Thus, the algorithm is suitable for simulating the complicated dissipation occurring in experimental systems.

\section{Statement of Simulation Problem} In general, classical computational algorithms for simulating a networks of oscillators have time complexity with polynomial dependence on $N$, the number of degrees of freedom \cite{wei-equilibriumstates-2009, rosa-dissipation-2008}. 
In the case of very large classical systems (i.e. $N = O(2^n)$), such as when one considers a large bath, these algorithms are 
not feasible. Ref. \cite{babbush-2023-oscillators}
gives a method for simulating $2^n$ classical coupled oscillators in $\poly(n)$ time on a quantum device. Moreover, they show that this problem is $\cB \cQ\cP$-Hard in addition to lying in $\cB \cQ\cP$, suggesting quantum advantage. 
However, their result assumes that the classical system is sparsely connected, so that each oscillator is coupled to at most $\poly(n)$ other oscillators. This is not the case in the CL model, where a single mass is connected to an exponential number of bath oscillators.

As shown in \cref{fig:system-fig}, we consider a finite system of $d$ oscillators coupled to a bath that is modeled by $N = 2^n$ independent degrees of freedom. We take $d \ll N$ and call such a system size \textit{efficient}. We refer to the $d$ degrees of freedom as the \emph{primary system} and the overall $d+N$ degrees of freedom as the \emph{composite system}. We emphasize that the specific couplings in the primary system have no sparsity constraints or similar restrictions. While the exponentially large nature of the bath makes it a candidate for obtaining quantum speedup, the couplings are no longer sparse and therefore the algorithm from \cite{babbush-2023-oscillators} loses its advantage. To overcome this issue, we rely on the internal structure of the graph corresponding to the adjacency matrix of the overall system, which can be exploited using quantum walks \cite{childs-2009-relationship, szegedy2004spectraquantizedwalkssqrtdeltaepsilon, childs-2010-limitations, childs-2012-blackbox}. 
Our result gives an $O(\poly(d, n, t, \eps^{-1}))$ algorithm for the following computational problem.

\begin{problem}\label{problem:main-problem}
Let $\clH$ be the classical Hamiltonian for a composite system, made up of $N = 2^n$ bath degrees of freedom, $\{y_{\alpha}\}_{\alpha \in [N]}$, with momenta $\{k_{\alpha}\}_{\alpha \in [N]}$, coupled to a single particle in a primary system of $d = \poly(n)$ oscillators, $\{x_i\}_{i \in [d]}$ with momenta $\{p_i\}_{i \in [d]}$, as defined in \cref{eq:multi-hamiltonian}. For the initial state specified by $\{x_i(0)\}$, $\{p_i(0)\}$, $\{y_{\alpha}(0)=(g_{\alpha}/\nu_{\alpha})x_{*}(0)\}$, $\{k_{\alpha}(0)=0\}$ and a given time $t$, output $\{x_i(t)\}, \{p_i(t)\}$ with precision $\eps$.
\end{problem}
\section{Modeling Dissipative Systems}\label{sec:modeling-dissipation}

The Caldeira-Leggett Hamiltonian, $\clH$, consists of two terms: the primary system Hamiltonian, $\clH_{\rm S}$, and the bath Hamiltonian, $\clH_{\rm B}$, defined as follows: 

\begin{subequations}\label{eq:multi-hamiltonian}\begin{gather}
\clH = \clH_{\rm S} + \clH_{\rm B}
\\ \clH_{\rm S} = \frac{1}{2}\sum_{i} \br{\frac{p_i^{2}}{m_i} + \kappa_{ii}x_i^{2}} + \frac{1}{2}\sum_{i\ne j}\kappa_{ij}(x_{i}-x_{j})^2
\\
\clH_{\rm B} = \frac{1}{2}\sum_{\alpha}\nu_{\alpha}\br{k_{\alpha}^{2} + \p{y_{\alpha} - \frac{g_{\alpha}}{\nu_{\alpha}}x_{*}}^{2}}
\end{gather}\end{subequations}

Here, $\{x_i,p_i\}_{i \in [d]}$ are the conjugate positions (defined as an offset from the system's equilibrium positions) and momenta of the primary degrees of freedom, while $\{y_{\alpha}, k_{\alpha}\}_{\alpha \in [N]}$ are those of the bath. We denote the position of the mass coupled to the bath by $x_*$. We let $\{m_i\}_{i \in [d]}$ be the masses of the primary system degrees of freedom and $\{\kappa_{ij}\}_{i,j \in [d]}$ the system's spring constants, such that $\kappa_{ij}=\kappa_{ji} \geq 0$ describes the coupling between the $i^{\mathsf{th}}$ and $j^{\mathsf{th}}$ oscillators and $\kappa_{ii}$ the coupling for the $i^\mathsf{th}$ oscillator to some wall. We let $ \{\nu_\alpha\}_{\alpha \in [N]}$ be the frequency of the $\alpha^{\mathsf{th}}$ bath oscillator, and $\{g_\alpha\}_{\alpha \in [N]}$ its coupling to mass $m_*$, where $m_*$ is the mass coupled to the bath.
For all $\alpha$, we take $g_\alpha\sim 1/\sqrt{N}$ to be small  and $0 \leq \nu_\alpha \leq \numax$ for some maximum frequency $\numax$.

The distribution of the frequencies $\nu_{\alpha}$ and couplings $g_{\alpha}$ determines the behavior of the bath. 
In the simplest case, one can take $N=1$; then the ``bath" simply adds one more natural frequency to system and shifts the existing ones.
In the other limit, Caldeira and Leggett take a constant coupling and a uniform frequency distribution stretching to infinity. One can then analytically recover that the primary system behaves according to $\clH_{\rm S}$ with the addition of dissipation and a white-noise fluctuation term, $\dot{x}_{*}\propto -\gamma x_{*} + \sqrt{f(t)}$. This is generally referred to as the Markovian limit, and the bath's time correlation structure is local, $\langle f(t)f(t^{\prime})\rangle \propto\delta(t-t^{\prime})$, meaning the it lacks of memory of previous events. The crossover between these limiting cases offers especially rich, interesting physics. In particular, taking a large bath with a frequency range  that is finite compared with the primary system's frequency, can lead to so-called non-Markovian dynamics, where the energy of the system dissipates but the bath has some finite-length memory and therefore non-trivial dynamics. This has been of particular interest for quantum systems \cite{devega-2017-dynamicsnonmarkov, gardiner-zoller-2004-quantum_noise, krovi2024quantumalgorithmssimulatequadratic}.

\Cref{eq:multi-hamiltonian} gives rise to a set of ordinary differential equations \cite{endmatter}. The task is to simulate these dynamics on a quantum computer, which we do by defining a quantum Hamiltonian that is suitable for Hamiltonian simulation. This Hamiltonian is directly derived from the equations of motion following the techniques of Ref.~\onlinecite{babbush-2023-oscillators}: 
\begin{equation}\label{eq:quantum-hamiltonian}
\quH = 
\begin{pmatrix}
    \mathbf{0} & i\sqrt{\matK}\sqrt{\matM}^{-1} & \mathbf{0} & \mathbf{0} \\
    -i\sqrt{\matM}^{-1}\sqrt{\matK} & \mathbf{0} & i\sqrt{\matM}^{-1}\matG & \mathbf{0}\\
    \mathbf{0} & -i\matG^{T}\sqrt{\matM}^{-1} & \mathbf{0} & i\matF \\
    \mathbf{0} & \mathbf{0} & -i\matF & \mathbf{0}
\end{pmatrix}
\end{equation}
Here, $\matM$ is the $d \times d$ diagonal matrix of masses, $\matM_{ij} = \delta_{ij}m_{i}$; 
$\matK$ is the positive-definite $d \times d$ matrix of spring constants, with $\matK_{ii} = \sum_{j}\kappa_{ij}$ and ${\matK_{ij} = -\kappa_{ij}}$ for $i\ne j$;
$\matF$ is the $N\times N$ diagonal matrix of bath frequencies, $\matF_{\alpha \beta} = \delta_{\alpha \beta}\nu_\alpha$; and $\matG$ is the $d \times N$ matrix with non-zero entries in a single row, $\matG_{i\alpha} = \delta_{i,{*}}g_{\alpha}/\sqrt{\nu_{\alpha}}$. (We have defined $\sqrt{\matA}$ as the principal square root of a positive semi-definite matrix $\matA$.)

If initialized at $t = 0$ in the following form and evolved under $\quH$, the quantum state remains related to the coordinates of the classical system by: 
\begin{equation}\label{eq:init-state}
    \ket{\psi(t)} = \frac{1}{\sqrt{2E_{0}}}
    \begin{pmatrix}
        \sqrt{\matK}\cdot \vecx(t)
        \\ \sqrt{\matM}^{-1}\cdot\vecp(t)
        \\ \sqrt{\matF}\cdot\vecy(t) - \matG^{T}\cdot \vecx(t)
        \\ \sqrt{\matF}\cdot\veck(t)
    \end{pmatrix}
\end{equation}
where $E_{0} = \clH$ is the total energy of the composite system, which is conserved over time. 

The final state $\ket{\psi(t)}$ can be used to extract $\{x_i(t)\}$ and $\{p_i(t)\}$ by performing state tomography within a subset of the Hilbert space containing the amplitudes proportional to the primary system coordinates. These consist of $2d$ states and so this process can be done with $\poly(d)$ repetitions \cite{1989-vogel-tomography, daniel-2001-tomography, knill-2007-measurement,Wong2025}.
\section{Hamiltonian Simulation}  
Hamiltonian simulation is the challenge of implementing the time evolution of a given quantum system on a quantum device. More precisely, it is the problem of constructing a unitary operator $\hat\matU\approx e^{-i \quH t} $ with precision $\eps$ that approximates the time evolution of a  Hamiltonian $\quH$ for a given time $t$. 
Determining $\hat\matU$ requires finding a representation of $\quH$ which can be compiled into a polynomial sequence of unitaries, and in turn, an efficient number of single and two-qubit quantum gates. Such a representation is straightforward in the case that $\quH$ is sparse (that is, contains at most $O(\poly(n))$ entries per row) \cite{low-qsp-2017, gilyen-2023-quantum}. 

In \cref{eq:quantum-hamiltonian}, $\quH$ is non-sparse, and in particular $\matG$ contains a row with $O(N = 2^n)$ terms. In this case techniques for simulation are less understood, the main issue being that a decomposition into simulable subsystems becomes significantly more challenging. 

Simulating the spectrum of the Hamiltonian, rather than $\quH$ directly, offers a partial solution, allowing for efficient simulation of $\quH$ in the case that its spectrum can be recovered using a quantum circuit \cite{childs-2009-relationship, childs-2010-limitations}. One uses the fact that the spectral decomposition of $\quH = \hat \matV \hat \matD \hat \matV^\dagger$ into a unitary $\hat\matV$ and diagonal $\hat\matD$ satisfies the following identity:
\begin{equation}\label{eq:spectrum-sim}
    e^{-i \hat\matV \hat\matD \hat\matV^\dagger t} = \hat\matV \cdot e^{-i \hat\matD t} \cdot \hat\matV^\dagger 
\end{equation}
Provided that one can efficiently implement $\hat\matV$ and apply the eigenvalues of $\quH$, simulation is straightforward. 

The work of \cite{childs-2009-relationship} uses this identity to develop a method for simulating arbitrary non-sparse Hamiltonians by combining quantum walks with phase estimation in a fixed basis and \cite{childs-2012-blackbox} implements this walk explicitly. However, as detailed in \cite{childs-2009-relationship}, the proposed method is limited in the sense that it requires preparing the principal eigenvector of $abs(\quH)$ (the element-wise absolute value of $\quH$ in a fixed basis) which generally is exponentially hard for Hamiltonians with negative or complex-valued entries, such as those of $\quH$ specified in \cref{eq:quantum-hamiltonian}. In \cite{childs-2010-limitations}, this issue is resolved by considering the graph obtained from the adjacency matrix defined by $\hat\matH$. Hamiltonians whose graphs are trees are shown to have $\|abs(\quH)\| = \| \quH \|$, allowing one to work with the basis independent quantity $\| \quH \|$ when simulating $\quH$. This result is also generalized to graphs with finite arboricity -- informally, the number of sets of trees that a given graph can be decomposed into \cite{eppstein-1994-arbor}. Namely, for graphs with arboricity $b$, it is shown in \cite{childs-2009-relationship} that $\|abs(\quH)\| \leq 2b \cdot \|\quH \|$, where $\| \cdot \|$ is the spectral norm, i.e.~the largest eigenvalue of $\quH$ by absolute value.

These bounds, combined with the algorithm from \cite{childs-2009-relationship,childs-2012-blackbox} result in an $O(b \cdot  \| \quH \| \cdot t \cdot \sqrt{\eps}^{-1})$ simulation for a non-sparse Hamiltonian $\quH$ with arboricity $b$. We note that measuring the complexity of simulation based on matrix norms is standard in the analysis of Hamiltonian simulation algorithms, as described in detail in \cite{childs-2010-limitations}. Since quantum phase estimation and quantum walks are both efficient quantum algorithms, the corresponding gate complexity is $O(\poly(b, n, t, \eps^{-1}))$ \cite{kitaev1995quantummeasurementsabelianstabilizer, Childs_2003}.
\section{Simulating our System}\label{sec:quantum-algorithm}

The method outlined above gives an efficient simulation algorithm for Hamiltonians whose graphs have small arboricity $b$ and small $\| \quH \|$. In the case of $\quH$ defined in \cref{eq:quantum-hamiltonian}, the spectrum of $\hat\matH$ is dominated by the entries of $\matF$, and therefore the greatest eigenvalue, $\lambda_{\max}$, of $\hat\matH$ is approximately $ \nu_{\max} = \max_k \nu_k$. It follows that $\| \quH \| \approx \numax$. 

\begin{figure}
\centering\includegraphics[width=\linewidth]{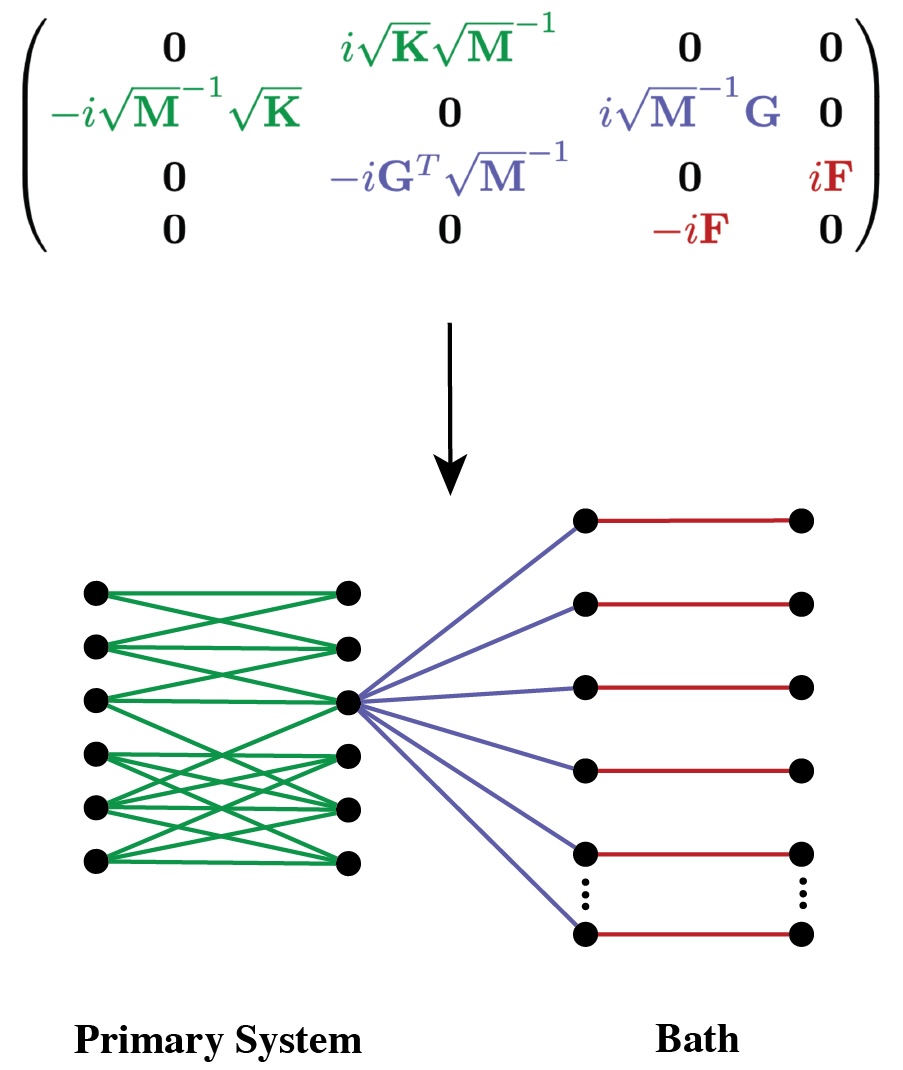}
    \caption{The graph of the Caldeira-Leggett model, obtained from the adjacency matrix specified by the system Hamiltonian.
    Each column of vertices corresponds to one of the rows or columns of the Hamiltonian of \cref{eq:quantum-hamiltonian}. 
    From right to left, we have $d$ vertices corresponding to the positions of the primary system; these are coupled by $\quH$ only to the $d$ primary momenta, in the second column. The momenta vertices are coupled only back to the primary positions, except for column $\rm s$, which is coupled by $\matG$ to the $N$ positions of the bath oscillators represented in the third column. Finally, each bath position vertex is coupled solely to the corresponding momentum by the diagonal $\matF$.
    }  \label{fig:hamiltonian-graph}
\end{figure}

We take $0 \leq \nu_k \leq \numax = \poly(n)$ and $g_k\sim 1/\sqrt{N}$ to be small for all $k$. Moreover, the graph of $\quH$ has arboricity $b \leq d$, as depicted in \cref{fig:hamiltonian-graph}. Therefore, the simulation algorithm runs in time $O(d \cdot \nu_{\max} \cdot t \cdot \sqrt{\eps}^{-1}) = O(d \cdot \poly(n) \cdot t \cdot \sqrt{\epsilon}^{-1})$ and we conclude that any efficient system size coupled to a bath may be simulated using the techniques from \cite{childs-2009-relationship, childs-2010-limitations, childs-2012-blackbox}.

Moreover, even for some exponentially large primary systems simulation is still be possible as long as the couplings in the primary system are sparse, i.e. for a system with $d = O(2^n)$ but $\matK$ that is $O(n)$-sparse. This can be done by combining the techniques outlined above with those in \cite{babbush-2023-oscillators}. Namely, the primary system can be simulated independently from the bath using sparse Hamiltonian simulation techniques while the bath interacting with the primary system degree of freedom, $x_{*}$, is simulated via the quantum walk approach outlined above, with $d = 1$. The two simulations may be combined using standard techniques \cite{childs-2012-blackbox}.

\section{Classical Hardness}\label{sec:computational-hardness}
The best known classical algorithms for simulating such a system scale polynomially in $N$ which leads to the possibility of exponential quantum advantage. While the algorithm in \cite{babbush-2023-oscillators} is shown to be $\cB \cQ\cP$-Hard,  a direct reduction is challenging because the process for converting an exponentially large graph that is sparsely connected to one that is dense is not efficient. However, we provide evidence for classical hardness by considering whether the quantum algorithm for simulating our system may be dequantized through the classical Quantum Singular Value transform (QSVT) \cite{bakshi-2023-improvedclassicalsingularvalue, gilyen-qsvt-2019, tang-2023-csguidequantumsingular}.

As described in \cite{bakshi-2023-improvedclassicalsingularvalue}, the corresponding classical algorithm would have dependence on the stable-rank of $\hat\matH$, which is defined as ${\|\hat\matH\|_F^2}/{\|\hat\matH\|^2}$, where $\| \cdot \|_F$ is the Frobenius norm. A short calculation shows that the stable-rank of $\hat\matH$ is $\exp(n)$ and thus the classical QSVT from \cite{bakshi-2023-improvedclassicalsingularvalue} is inefficient \cite{supplemental}, suggesting that the problem we address in this work has no efficient classical counterparts. 
\section{Discussion}
In this work, we present for the first time an efficient quantum algorithm for simulating a quadratic open system coupled to an exponentially large bath. We achieve this by giving an explicit reduction from the Caldeira-Leggett model to a quantum Hamiltonian simulation. The algorithm is based on discrete-time quantum walks for implementing non-sparse Hamiltonian simulation. This could be used to provide speedups for various settings of classical particle dynamics \cite{moeendarbary-dpd-2009, Li-dpd-2016}. 
The methods in this work could possibly extend to the quantum regime, as well. The techniques shown here are already sufficient for systems with quadratic Hamiltonians, and it is possible that they may be extended to the case of a non-linear primary system coupled to some Gaussian bath.

\section{Acknowledgements}
\'{A}gi Vill\'{a}nyi acknowledges support by the Doc Bedard fellowship from the Laboratory for Physical Sciences through the MIT Center for Quantum Engineering and the National Science Foundation Graduate Research Fellowship under Grant No. 2141064. 

\bibliography{quantum}

\clearpage
\appendix
\newpage
\onecolumngrid
\begin{center}
{\textbf{End Matter}}
\end{center}
\twocolumngrid
\section{Modeling Dissipation} The equations of motion corresponding to the Caldeira-Leggett Hamiltonian defined in \cref{eq:multi-hamiltonian} are: 

\begin{subequations}\label{eq:equs-of-motion}
\begin{align}
\dot{x}_i &= p_i/m_i
\\ \begin{split}\dot{p}_i &= -\kappa_{ii}x_i - \sum_{j \neq i}\kappa_{ij}(x_i - x_j) \\ &\quad + \delta_{i,*} \sum_{\alpha} g_{\alpha} \p{y_{\alpha} - \frac{g_{\alpha}}{\nu_{\alpha}}x_{*}}
\end{split}
\\ \dot{y}_{\alpha} &= \nu_{\alpha}k_{\alpha} 
\\ \dot{k}_{\alpha} &= -\nu_{\alpha}y_{\alpha} + g_{\alpha}x_{*}
\end{align}\end{subequations}
Below we demonstrate that the Hamiltonian of the composite system defined in \cref{eq:quantum-hamiltonian} satisfies these equations.

In matrix form: 
\begin{subequations}\label{eq:equs-of-motion-matrix}
\begin{align}
\dot{\vecx}(t) &= {\matM}^{-1} \cdot \vecp(t)
\\ \dot{\vecp}(t) &= -\matK \cdot \vecx(t) + \matG\left( \sqrt{\matF}\cdot \vecy(t) -  \matG^T \cdot \vecx(t)\right)
\\ \dot{\vecy}(t) &= \matF \cdot \veck(t)
\\ \dot{\veck}(t) &= -\matF \cdot \vecy(t) + 
\sqrt{\matF} \matG^T \cdot \vecx(t)
\end{align}\end{subequations}
Then, from 
\begin{equation*}
\frac{d}{dt}\ket{\psi(t)} = -i\hat\matH\ket{\psi(t)},
\end{equation*}
(with $\hbar = 1$) we get the equations of motion, with initial state $\ket{\psi(t)}$ as defined in \cref{eq:init-state}. Namely, writing out $\frac{d}{dt}\ket{\psi(t)}$ explicitly: 

\begin{widetext}
\begin{equation*}\label{eq:lhs}
 \p{\begin{array}{c}
 \sqrt{\matK}\cdot \dot{\vecx}(t)
        \\ \sqrt{\matM}^{-1}\cdot\dot{\vecp}(t)
        \\ \sqrt{\matF}\cdot\dot{\vecy}(t) - \matG^{T}\cdot \dot{\vecx}(t)
        \\ \sqrt{\matF}\cdot \dot{\veck}(t)
\end{array}}=
\p{\begin{array}{c}
\sqrt{\matK} {\matM}^{-1}\cdot\vecp(t) \\
\sqrt{\matM}^{-1}\left[-\matK \cdot  \vecx(t) + \matG\left(\sqrt{\matF}\cdot \vecy(t) -  \matG^T \cdot \vecx(t)\right)\right] \\ 
\sqrt{\matF} \matF \cdot \veck(t) - \matG^T\matM^{-1}\cdot \vecp(t) \\
-\sqrt{\matF}\matF \cdot \vecy(t) + \matF \matG^T \cdot \vecx(t) 
\end{array}}.
\end{equation*}
This matches the result of $-i \hat\matH \ket{\psi(t)}$. 
\end{widetext}

\section{Dequantization} Below we detail the calculation for the stable-rank of $\hat\matH$, where the stable-rank is given by ${\|\hat\matH\|_F^2}/{\|\hat\matH\|^2}$. The Frobenius norm of $\hat\matH$ is
\begin{equation}
\begin{aligned}
     & \|\quH\|_F^2 =  \sum_{\ell,\ell^{\prime}=1}^{2(d+N)} |\quH(\ell, \ell^{\prime})|^2
\end{aligned}
\end{equation}
and substituting the definitions above 
\begin{equation}
    \begin{aligned}
        \|\quH\|_F^2 &= 2\left(\|\sqrt{\matK}\sqrt{\matM}^{-1}\|_F^2 + \|\sqrt{\matM}^{-1}\matG\|_F^2 + \|\matF\|_F^2\right)
        \\ & = 
        2 \left[\sum_{i}^{d} \frac{\matK_{ii}}{m_{i}}
        + \sum_{\alpha=1}^{N}\frac{g_{\alpha}^2}{m_{*}\nu_{\alpha}} + \sum_{\alpha=1}^{N}\nu_{\alpha}^2\right]
    \end{aligned}
\end{equation}

Meanwhile, the spectral norm has
\begin{equation}
    \begin{aligned}
        \|\quH\| &= \max_{\vecv \in \mathbb{C}^n, \|\vecv\| = 1} \|\quH\vecv\|_2\\
        &= (\max_i |\lambda_i |) \\
        &\approx (\max_{\alpha} \nu_{\alpha}) \quad \text{since } g_{\alpha} \ll \nu_{\alpha} \text{ } \forall {\alpha}
    \end{aligned}
\end{equation}

Therefore, ${\|\hat\matH\|_F^2}/{\|\hat\matH\|^2}$ is $\exp(n)$ and the classical algorithm from \cite{bakshi-2023-improvedclassicalsingularvalue} is inefficient.

\section{Details of Algorithm} Below we give more details about our algorithm which combines results from Section 5 of \cite{childs-2009-relationship}, and Propositions 1 and 2 in \cite{childs-2010-limitations}. 

Recall that we are simulating the $D \times D$ Hamiltonian $\hat\matH$, where $D = 2(d + N)$. By Proposition 1 in \cite{childs-2010-limitations}, $\|abs(\quH)\| = \| \quH \|$ when the graph of $\quH$ is a tree, which applies to this work. For clarity of presentation, we recall the algorithm from \cite{childs-2009-relationship} for $d=1$ for the special case $\|abs(\quH)\| = \|\quH\|$, followed by a brief discussion of the case of $d \geq 2$. 

The algorithm simulates $\quH$ by simulating its spectrum. It takes as input a state $\ket{\phi_0}$ in the eigenbasis of $\quH/\|\quH\|$. Namely, for eigenvectors $\{\ket{\ell}\}$ and corresponding eigenvalues $\gamma_\ell$: 

\begin{equation}
    \ket{\phi_0} = \sum_{\ell=1}^D \gamma_\ell \ket{\ell}
\end{equation}

The goal of the algorithm is to induce the phase in  $e^{-i \cdot \ell  \cdot t}$ for each term in $\ket{\phi_0}$, which simulates $\quH$ by \Cref{eq:spectrum-sim}. This requires performing phase estimation on some object that is related to the spectrum of $\quH$; namely, a quantum walk. 

The algorithm first maps the above state to the basis of the quantum walk, which is achieved using the isometry $\matT$ defined as follows: 

\begin{equation}
    \matT = \sum_{\alpha = 1}^D \ket{\psi_\alpha}\bra{\alpha}
\end{equation}
    
Here, $\ket{\psi_\alpha} \in \mathbb{C}^D \otimes \mathbb{C}^D$ depends on the entries of $\hat\matH^*_{\alpha\beta}$ and the coefficients of $\ket{v} = \sum_{i=1}^D v_i \ket{i}$, which is the eigenvector of $abs(\hat\matH)$ with eigenvalue $\|\hat\matH\|$:

\begin{equation}
    \ket{\psi_\alpha} = \|\hat\matH\|^{-1}\sum_{\beta=1}^N \left({\hat\matH_{\alpha\beta}^* \frac{v_\beta}{v_\alpha}}\right)^{-1/2}\ket{\alpha}\ket{\beta}
\end{equation}

As discussed in Section 3 of \cite{childs-2009-relationship}, in order for the simulation algorithm to be efficient, it is required that $\matT$ may be implemented efficiently, which means computing $v_\beta/ v_\alpha$ must be possible. While in general computing $\ket{v}$ is difficult, it is possible to compute $\ket{v}$ explicitly in special cases such as when the graph of the Hamiltonian is a tree, which is the case that is relevant to our work \cite{childs-2009-relationship, childs-2010-limitations}.

The state $\ket{\psi_\alpha}$ specifies the elements of a quantum walk. Namely, the walk is over the edges of the graph (unlike the case in classical random walks), which is required to make the mapping unitary \cite{szegedy2004spectraquantizedwalkssqrtdeltaepsilon, Childs_2003}. A single iteration of the quantum walk is obtained by first reflecting over the subspace spanned by $\{\ket{\psi_\alpha}\}$ and then exchanging registers $\ket{\alpha}\ket{\beta}$ by a standard swap operator $\matS$. The isometry $\matT$ given above defines the projector onto the subspace spanned by $\{\ket{\psi_\alpha}\}$, which is $\matT\matT^\dagger$. Therefore, the walk operator is $\matW = \matS\cdot (2\matT\matT^\dagger - \matI)$. The state $\ket{\phi_1} \in \mathbb{C}^D \otimes \mathbb{C}^D$ resulting from executing $\matT \ket{\phi_0}$ is: 

\begin{equation}
    \ket{\phi_1} = \sum_\ell \gamma_\ell \cdot \matT \ket{\ell}
\end{equation}

As detailed in \cite{childs-2009-relationship}, the state $\ket{\phi_1} \in \mathbb{C}^D \otimes \mathbb{C}^D$ resulting from executing $\matT \ket{\phi_0}$ may be written in terms of the spectrum of $\matW$. Namely:

\begin{equation}
    \ket{\phi_1} = \sum_\ell \gamma_\ell \left( \eta \ket{w_+} - \eta^* \ket{w_-} \right)
\end{equation}

Here, we take $s_\ell = sin^{-1}(\ell)$ and $c_\ell = cos^{-1}(\ell)$ as shorthand and we let $\eta = (1-\ell \cdot e^{-i \cdot c_\ell})/\sqrt{2(1-\ell^2)}$. The state $\ket{w_{\pm}}$ corresponds to the eigenvectors of $\matW$ with eigenvalues $w_{\pm} = \pm e^{\pm i \cdot s_\ell}$.

It remains to discuss how to extract the phase $e^{-i \cdot \ell \cdot t}$ by phase estimation. We only give an overview here, Section 5 of \cite{childs-2009-relationship} gives detailed analysis.  Performing the phase estimation procedure, $\matQ$, on $\ket{\phi_1}$ with query access to $\matW$ appends the result of the phase estimation, $\tilde{s}_\ell \approx s_\ell$, to an ancilla register. A unitary operation $\matP$ is then executed to apply the phase $e^{-i \cdot t \cdot sin(\tilde{s}_\ell)} = e^{-i \cdot t \cdot \tilde{\ell}}$, followed by uncomputing $\matQ$ and $\matT$ to return the system to the desired final state. In summary, in the case of a perfect phase estimation, the simulation algorithm is specified by the unitary: 
\begin{equation}
    \matU = \matT^\dagger \matQ^\dagger \matP \matQ \matT
\end{equation}

The runtime of the simulation is determined by the normalization parameter $\| \quH \|$ and the number of repetitions required to make the phase estimation step robust. As shown in Section 5 of \cite{childs-2009-relationship}, $O(t \cdot  \sqrt{\eps}^{-1})$ repetitions suffice. Therefore, the total runtime of the simulation is $O(\| \quH \| \cdot t \cdot \sqrt{\eps}^{-1})$. Since $\| \quH \| \approx \nu_{\max}$, the total runtime is $O(\nu_{\max} \cdot t \cdot \sqrt{\eps}^{-1})$ in the case of a singular primary system degree of freedom.

By Proposition 2 in \cite{childs-2010-limitations}, the runtime depends on the arboricity of the graph corresponding to $\quH$ for $d \geq 2$. In the case of $\quH$ considered in this work, the arboricity is $O(d)$, and therefore the total runtime is $O(d \cdot \nu_{\max} \cdot t \cdot \sqrt{\eps}^{-1})$. Since we take $\nu_{\max} = \poly(n)$, this is equivalent to $O(d \cdot \poly(n) \cdot t \cdot \sqrt{\eps}^{-1})$.

\end{document}